\documentclass[journal]{IEEEtran}
\IEEEoverridecommandlockouts
\usepackage{cite}
\usepackage{amsmath,amssymb,amsfonts}
\usepackage{algorithmic}
\usepackage{commath}
\usepackage{graphicx}
\usepackage{textcomp}
\usepackage[table,xcdraw]{xcolor}
\def\BibTeX{{\rm B\kern-.05em{\sc i\kern-.025em b}\kern-.08em
    T\kern-.1667em\lower.7ex\hbox{E}\kern-.125emX}}

\begin{document}
\bstctlcite{IEEEexample:BSTcontrol}

\title{Energy Efficient In-memory Hyperdimensional Encoding for Spatio-temporal Signal Processing
}
\author{\IEEEauthorblockN{Geethan Karunaratne\IEEEauthorrefmark{1}\IEEEauthorrefmark{3},
Manuel Le Gallo\IEEEauthorrefmark{1},
Michael Hersche\IEEEauthorrefmark{1}\IEEEauthorrefmark{3}, \\
Giovanni Cherubini\IEEEauthorrefmark{1},
Luca Benini\IEEEauthorrefmark{3},
Abu Sebastian\IEEEauthorrefmark{1}, and
Abbas Rahimi\IEEEauthorrefmark{1}}\\
\IEEEauthorblockA{\IEEEauthorrefmark{1}IBM Research,
Z\"{u}rich, Switzerland 
\IEEEauthorrefmark{3}ETH Z\"{u}rich, Z\"{u}rich, Switzerland}
}
\IEEEoverridecommandlockouts
\IEEEpubid{\parbox[t]{\columnwidth}{\vspace{-19mm}\textcopyright 2021 IEEE.  Personal use of this material is permitted. Permission from IEEE must be obtained for all other uses, in any current or future media, including reprinting/republishing this material for advertising or promotional purposes, creating new collective works, for resale or redistribution to servers or lists, or reuse of any copyrighted component of this work in other works. \newline
IEEE Transactions on Circuits and Systems II: Express Briefs\\
DOI: 10.1109/TCSII.2021.3068126\\URL: https://ieeexplore.ieee.org/document/9387248\hfill} \hspace{\columnsep}\makebox[\columnwidth]{ }}

\maketitle
\IEEEpubidadjcol
\begin{abstract}
The emerging brain-inspired computing paradigm known as hyperdimensional computing (HDC) has been proven to provide a lightweight learning framework for various cognitive tasks compared to the widely used deep learning-based approaches. Spatio-temporal (ST) signal processing, which encompasses biosignals such as electromyography (EMG) and electroencephalography (EEG), is one family of applications that could benefit from an HDC-based learning framework. At the core of HDC lie manipulations and comparisons of large bit patterns, which are inherently ill-suited to conventional computing platforms based on the von-Neumann architecture. In this work, we propose an architecture for ST signal processing within the HDC framework using predominantly in-memory compute arrays. In particular, we introduce a methodology for the in-memory hyperdimensional encoding of ST data to be used together with an in-memory associative search module. We show that the in-memory HDC encoder for ST signals offers at least 1.80$\times$ energy efficiency gains, 3.36$\times$ area gains, as well as 9.74$\times$ throughput gains compared with a dedicated digital hardware implementation. At the same time it achieves a peak classification accuracy within 0.04\% of that of the baseline HDC framework.
\end{abstract}


\section{Introduction}
\IEEEPARstart{A}{lmost} all breakthroughs in artificial intelligence in the last decade are characterized by an underlying machine learning model that entails a higher complexity in terms of the number of operations and parameters compared to contemporary models. 
Such increased model complexity demands more energy to perform training and inference tasks. 
Nevertheless, the human brain, with its far-reaching cognitive capabilities, consumes several orders of magnitude less power. 
This disparity has paved the way to exploring brain-inspired alternatives. Hyperdimensional computing (HDC)~\cite{Kanerva2009} is one promising brain-inspired computing approach that relies on representing entities using high-dimensional (up to 10,000 dimensions) vectors called \textit{hypervectors}. Similarly to the brain, where representations are spread across thousands of randomly originated neurons, a set of (pseudo)random orthogonal hypervectors forms the basis in the HDC framework. These hypervectors are then combined and compared using a well-defined set of algebraic operations to derive representations \newpage\noindent for composite entities and to find similarities, respectively.

HDC has been deployed in a diverse set of application domains, for instance in solving Raven’s progressive matrices~\cite{Raven}, analogical reasoning~\cite{Dollar-Mexico},
natural language processing~\cite{hyperembed_nlp}, robotics~\cite{sensorimotor_hd,dvs_hd}, text classification~\cite{Language_HD_17,Rahimi_lang_recog,HD_news,Karunaratne2020}, 
activity recognition \cite{activity_hetero_2014}, DNA sequencing~\cite{hdna}, and biosignal processing~\cite{rahimi_emg,emgmoin,pulphd,Rahimi_eeg2,HD_iEEG,Moin2021} (see~\cite{HD_ExG} for an overview). 

A promising application domain for HDC is the spatio-temporal (ST) processing of signals, acquired by EMG sensors for example. 
The corresponding HDC algorithm was presented in~\cite{rahimi_emg}, and later scaled up for high-density flexible EMG sensors~\cite{emgmoin,Moin2021}.
The same HDC algorithm has been used in a variety of applications such as EEG~\cite{Rahimi_eeg2}, iEEG~\cite{HD_iEEG}, ExG~\cite{HD_ExG} in general, as well as speech recognition~\cite{voicehd}, delivering higher classification accuracy than the established approaches. 
ST signal processing differs from other classes of applications because numerical data sequences received from multiple channels within a certain time window are considered as the input. 
Due to often being deployed at the edge of the Internet of Things and the confidential nature of the input data, the ST applications identified above could benefit immensely from energy-efficient hardware platforms. 

Earlier works address ST HDC signal processing in low-power hardware platforms, such as \textit{PULP}~\cite{pulphd} and \textit{ARM Cortex-A53}~\cite{activity_hd}. 
For example, PULP-HD~\cite{pulphd} describes the implementation of the ST HD encoder on multiple cores in a PULP cluster for achieving a target of 10\,ms detection latency in real-time.
Nevertheless, the energy efficiency could be improved further by using in-memory computing approaches~ \cite{VLSI_RRAM,cnfet_rram_HD,Karunaratne2020}. 
In-memory computing is an emerging paradigm where the physical attributes of memory devices are exploited to compute in place~\cite{Y2020sebastianNatNano}. Operations that require manipulation and comparison of large strings of bit patterns, which are at the core of the HDC framework, are particularly well suited for in-memory computing~\cite{Karunaratne2020}.

\begin{figure*}[t]
        \centering \includegraphics[width=\linewidth]{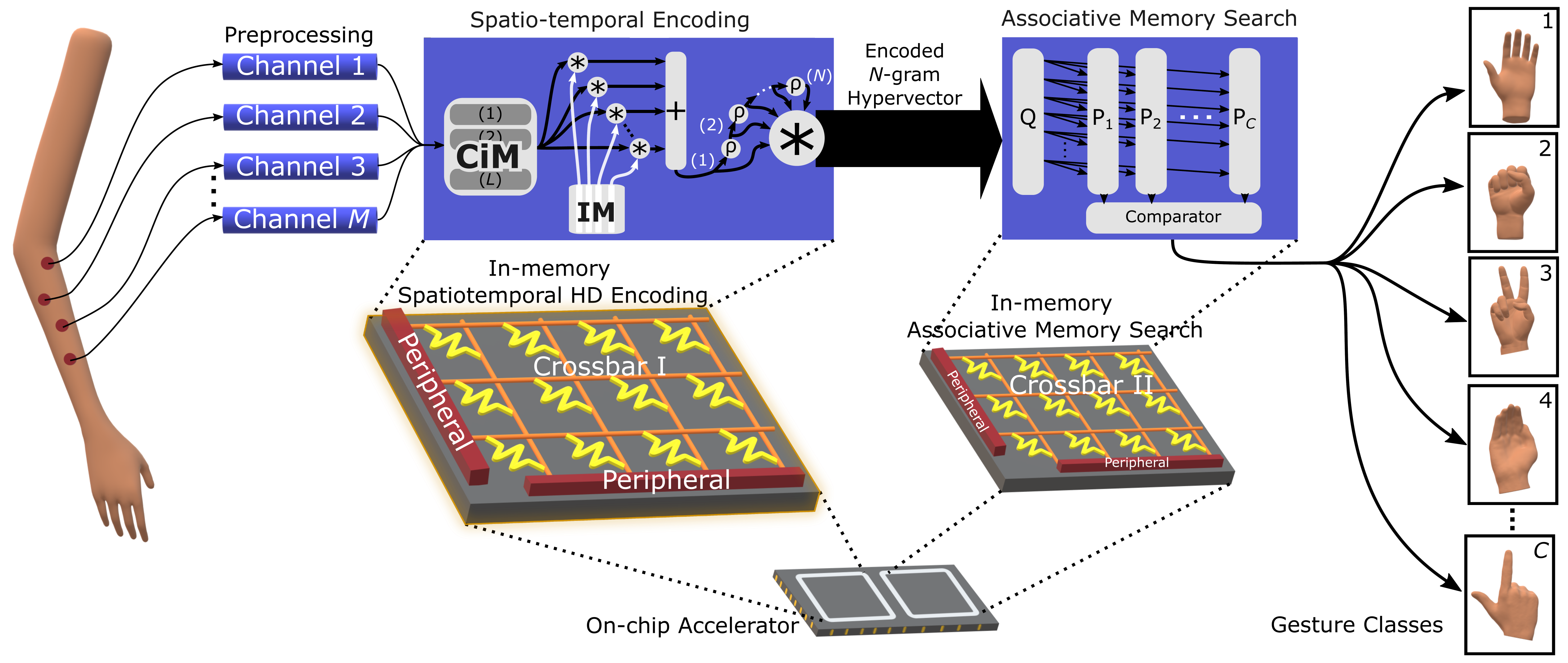}
        \caption{Concept of in-memory hyperdimensional encoding of spatio-temporal signals within the application of hand gesture recognition. First, the EMG signals are acquired from the electrodes connected to different parts of the subject's arm. After a pre-processing step, the data from different channels are embedded into one hypervector using spatial encoding and temporal encoding in a first memristive crossbar array, which is the focus of this work. The resulting query vector is passed to a second crossbar array to perform the associative memory search. The final result collected from the peripheral of the second crossbar predicts the class of the hand gesture.}
          \label{fig:concept_figure}
\end{figure*}

In this work, we propose an in-memory computing-based system for ST signal processing with HDC. An illustration of the system is given in Fig.~\ref{fig:concept_figure}, taking the EMG-based hand gesture recognition as a use case. Compared to prior work on in-memory HDC encoding~\cite{Karunaratne2020}, we present a novel in-memory computing HDC encoding architecture tailored for ST inputs. When coupled with an in-memory associative memory search module, such as the one presented in \cite{Karunaratne2020}, we get a complete in-memory HDC processor for ST signal processing. We derive classification accuracy results from simulations using a statistical model of phase-change memory (PCM) crossbar arrays. Furthermore, we estimate the throughput, area, and energy efficiency of the in-memory ST HDC encoder and compare it with a dedicated digital encoder as well as a ST HDC encoder running on a low-power general purpose compute platform, PULP-HD~\cite{pulphd}. 



\section{Algorithm}
\label{sec:algo}
\subsection{Conventional ST HDC encoding algorithm}
In the conventional ST HDC encoding algorithm \cite{rahimi_emg}, the input acquired from each channel goes through several pre-processing steps \cite{pulphd} and is converted to a stream of discrete time samples.
An HDC encoder maintains a record of \textit{N} consecutive time samples from \textit{M} channels to generate an encoded \textit{N}-gram hypervector embedding defined as:
\begin{equation}
f: \{{s}_{n,m}\}^{\textit{N}\times \textit{M}} \in \{l_1,l_2,...,l_L\}^{\textit{N}\times \textit{M}} \rightarrow G\in \{0,1\}^\textit{D},\label{eq:enc}
\end{equation}
where $n\in \{1,2,...,\textit{N}\}$ is the relative time index, $m\in \{1,2,...,\textit{M}\}$ denotes the source channel index and $\{l_1,l_2,...,l_L\}$ are the \textit{L} discrete quantization levels given in ascending order $l_1<l_2<...<l_L$. The output of the embedding function $G$ is a \textit{$\textit{D}$}-dimensional binary hypervector.

First, the encoder projects the sample values $s_{n,m}$ to a high dimensional space using a so-called continuous item memory ($CiM$)~\cite{rahimi_emg}. 
As opposed to assigning quasi-orthogonal hypervectors to every unique discrete sample value, the $CiM$ sets the Hamming distance ($HamD$) between two vectors to be proportional to the absolute difference between the corresponding sample values.
This is achieved by choosing quasi-orthogonal hypervectors for the minimum and maximum levels $l_1$ and $l_L$, and hypervectors corresponding to the intermediate levels that satisfy:
\begin{equation}
HamD(CiM(l_i),CiM(l_j)) = Floor\left(\frac{\textit{D}\abs{l_i-l_j}}{2\abs{l_L-l_1}}\right)\label{eq:cim}
\end{equation}

The $CiM$-projected vectors are then bound to the relevant channel ID hypervector $E_m$. The channel ID hypervectors are stored in an item memory ($IM$) and are quasi-orthogonal. The channel-bound hypervectors are given by $I_{m,n} = CiM(s_{n,m})*E_m$, where $*$ denotes the element-wise XOR binding operation. The channel-bound hypervectors are then bundled together to produce spatial hypervectors 
\begin{equation}\label{eq:majority_1}
S_n = Majority(I_{1,n},I_{2,n},...,I_{\textit{M},n}),
\end{equation}
where, for each dimension, the majority function outputs \textit{1}(\textit{0}) if the majority of channels have \textit{1}(\textit{0}). 

The spatial hypervectors $S_1,S_2,...,S_N$ then enter a temporal encoding stage, which outputs the final encoded \textit{N}-gram $G$ according to the following equation:
\begin{equation}
G_t = \rho^{\textit{N}-1}S_1*\rho^{\textit{N}-2}S_2*...*S_N\label{eq:temporal_1}
\end{equation}
where $t$ is the time step at which the $\textit{N}^{th}$ sample of the current data record enters the system, and $\rho$ is the vector permutation operator, which is implemented as a circular right shift.

During the training phase, \textit{N}-grams collected from the same class are further bundled to produce \textit{class prototype hypervectors} $P_c$, where $c\in\{1,2,...,C\}$, and $C$ is the number of classes. During the inference phase, the \textit{N}-gram produced from the same encoder is called a \textit{query hypervector} $Q$ and used to measure binary dot product similarity against each of the prototype hypervectors $P_c$. The class with the highest similarity is selected as the predicted class.

\subsection{Adaptations to suit in-memory computing}
\label{sec:im_algo}
We propose several adaptations to the conventional algorithm to suit in-memory computing. First, all channel-bound hypervectors are pre-computed and unrolled using:
\begin{equation}
I^l_m = CiM(l)*E_m 	\quad\forall l\in\{1,...,\textit{L}\}, \forall m\in\{1,...,\textit{M}\}\label{eq:channelbound}
\end{equation}
and stored in the memristive crossbar array. The other major modification to the conventional ST encoding algorithm is that the temporal encoding is performed immediately on each of the channel-bound hypervectors to obtain intermediate temporal hypervectors $T_m$ as given in:
\begin{equation}
T_m = \rho^{\textit{N}-1}I_{1,m}*\rho^{\textit{N}-2}I_{2,m}*...*I_{\textit{N},m}\label{eq:temporal_2}
\end{equation}
instead of waiting for the spatial bundling given in \eqref{eq:majority_1}. Finally, the hypervectors $T_m$ are bundled to produce the adapted \textit{N}-gram hypervector $G'$, given by:
\begin{equation}
G'_t = Majority(T_{1},T_{2},...,T_{\textit{M}})\label{eq:majority_2}
\end{equation}
In summary, compared to the ST encoder in \cite{rahimi_emg}, the proposed in-memory ST encoding algorithm pre-computes channel-bound hypervectors and pushes the bundling portion of the spatial encoding step downstream of the temporal encoding step. This offers further  flexibility to set individual quantization levels and \textit{N}-gram sizes per channel, which is not possible with the conventional encoder. This is a useful feature to exploit channel specific spatial and temporal dynamics. 

\section{Architecture}
\label{sec:Archi}
\begin{figure}[tb]
\centering
\includegraphics[width=1\columnwidth]{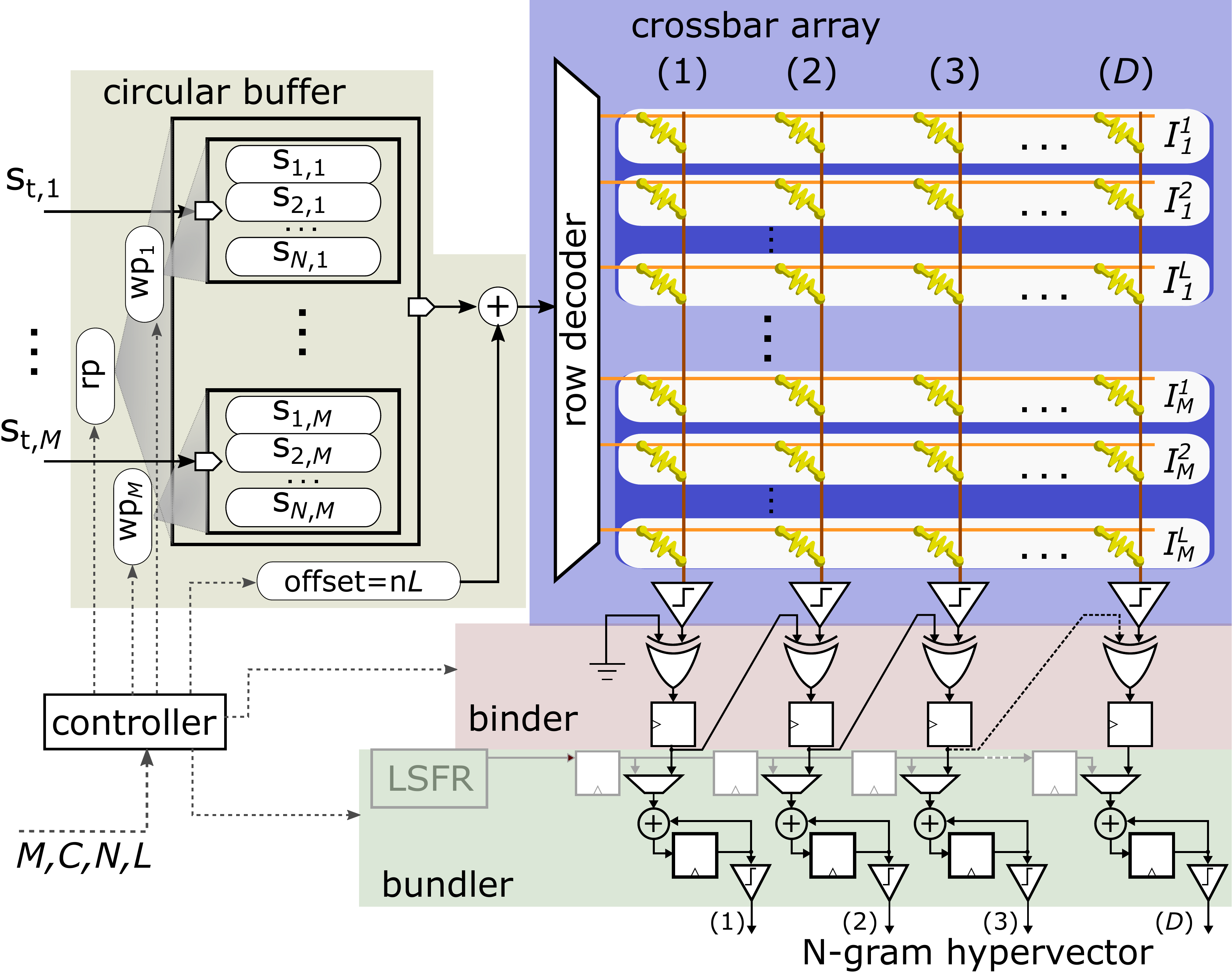}
\caption{In-memory spatio-temporal HDC encoding architecture.}
\label{fig:arch}
\end{figure}
The architecture of the in-memory ST HDC encoder is shown in Fig.~\ref{fig:arch}. It consists of a memristive crossbar array and a few peripheral circuits, namely a circular buffer, a binder, and a bundler. The circular buffer is maintains the last \textit{N} samples and reading them sequentially. It has $\textit{M}$ write pointers ($wp_1$, ..., $wp_\textit{M}$), synchronized to each one of the low-frequency external input channels that write data in parallel to the next allotted locations in the buffer. There is a single read pointer that is synchronized to the internal clock frequency which is set at least $NM\times$ faster than the external frequency to avoid any data loss. The read pointer ($rp$) traverses the whole input data record sequentially: it samples in chronological order within each channel and repeats over all channels.

As shown in Fig.~\ref{fig:arch}, the pre-computed channel-bound hypervectors $I_{m}^l$ given in \eqref{eq:channelbound} are stored along the rows of the $\textit{M}\!\cdot\! \textit{L}\times \textit{D}$ crossbar array. 
This allows us to save $\textit{D}\!\cdot\! \textit{N}\!\cdot\! M$ XOR operations per input data record. 
Performing temporal encoding on each channel separately allows us to reduce the number of intermediate buffers in the digital domain that must possess read/write capability at the expense of additional read-only storage in the PCM crossbar array. This is an acceptable trade-off because the PCM device consumes approximately 23\,fJ of energy per read operation \cite{mixed_precision_PCM}. This is just a fraction of the energy incurred by digital read/write buffers. Furthermore, thanks to their non-volatile nature, PCM devices do not consume energy when retaining their content in idle mode.

The output of the crossbar array is connected to an array of sense amplifiers. Sequential processing of the data record allows time sharing the sense amplifier array and the downstream binder module, allowing us to save a significant amount of energy and area in the peripherals. The binder module consists of an array of $\textit{D}$ XOR gates daisy-chained with $\textit{D}$ registers, which collectively implement \eqref{eq:temporal_2}.

The bundler module in the architecture implements \eqref{eq:majority_2}. An optional scan chain, which propagates a random hypervector generated bit-by-bit from a linear feedback shift register, is activated at the start of the encoding cycle when the number of input channels is even, with ties that are broken randomly. The majority function is implemented as an array of $log_2\textit{M}$-bit accumulators, followed by an array of comparators whose reference is set at $ceil((\textit{M}+1)/2)-0.5$. 

The controller module receives the encoding parameters and coordinates the flow across the rest of the modules. For example, it communicates the start/end addresses for each of the write pointers, the offset value added to circular buffer read data to derive the row address in the crossbar array, when to update the 1-bit register array in the binder module, when to update accumulators in the bundler, etc.

\section{Results and Discussion}
\label{sec:Results}
\subsection{Experimental setup}
The proposed in-memory ST HDC architecture is benchmarked on the EMG hand gesture recognition dataset \cite{rahimi_emg}. It includes data acquired from five subjects who perform five classes of gestures. The data is sampled at a 500\;Hz frequency  via four EMG electrodes attached to each subject's forearm. The class label and channel readings are provided at each time frame. We use 25\% of the 175$\times$ down-sampled data to train a 10,000-\textit{D} HDC model for each subject and test on 800 queries on average per subject. The result is averaged across the subjects to obtain the classification accuracy. For PCM simulations, the statistical model described in \cite{Karunaratne2020} is used, which captures non-ideal effects such as spatial and temporal variations in the PCM crossbar array.

\subsection{Classification accuracy results}
\label{sec:accuracy}
\begin{figure}[tb]
\centering
\includegraphics[width=1\columnwidth]{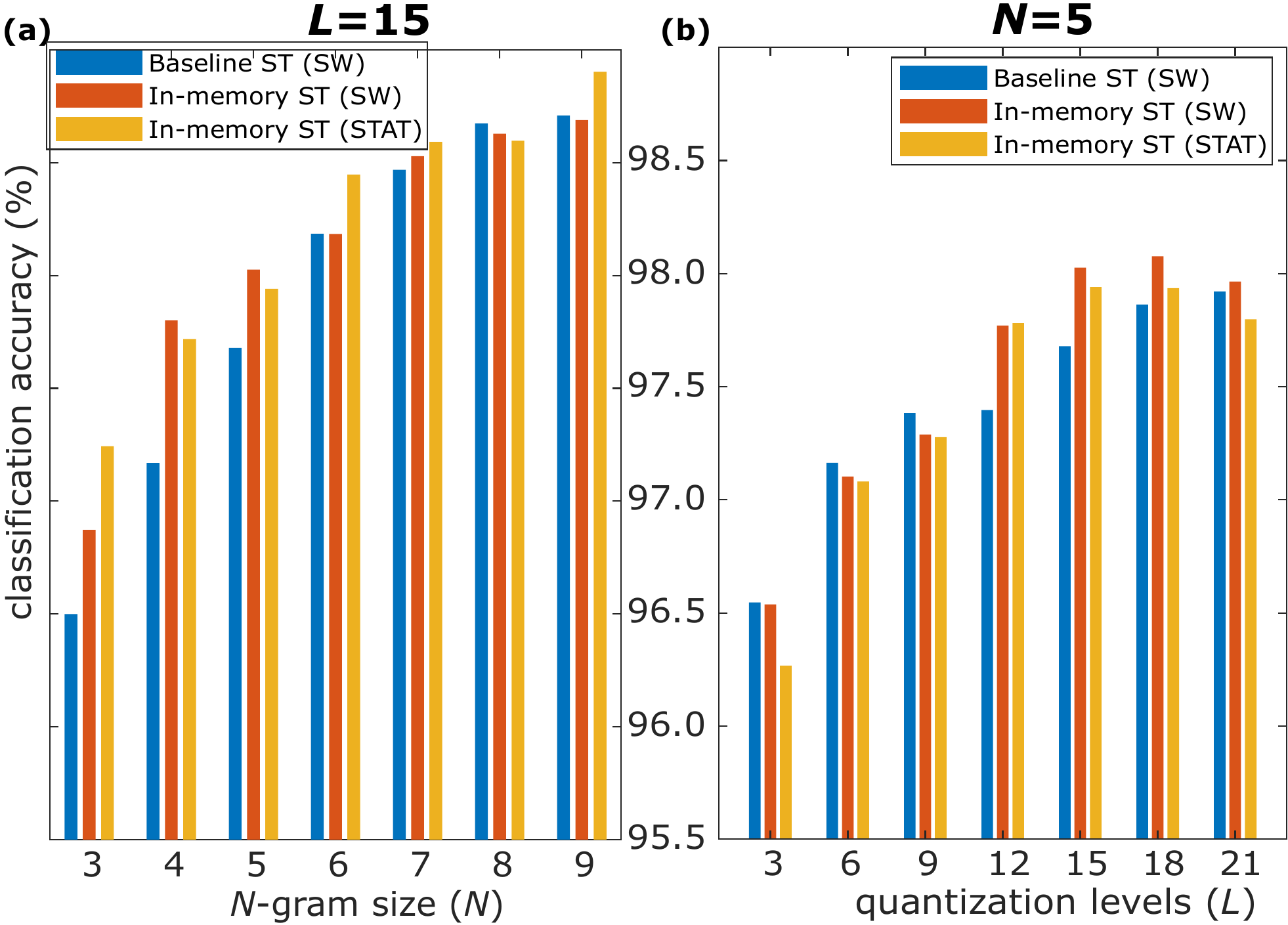}
\caption{Classification accuracies obtained from a baseline software ST encoder, an in-memory friendly ST encoder running in software, and an in-memory ST encoder simulated using the PCM statistical model. In (a), accuracy variation with \textit{N}-gram size for a fixed quantization level is plotted. In (b), accuracy variation with quantization levels for a fixed \textit{N}-gram size is plotted.}
\label{fig:accuracy}
\end{figure}

Fig.~\ref{fig:accuracy} shows the classification accuracy obtained from an in-memory ST encoder running in software, as well as the same encoder simulated using the statistical model, and comparing these results with the baseline ST encoder \cite{rahimi_emg}. In all three models of encoders, the in-memory associative memory search module is also simulated with the statistical model.
The proposed in-memory ST encoder achieves a peak accuracy of 98.9\% (see Fig.~\ref{fig:accuracy}(a)) when {\textit{N}=9, \textit{L}=15}. This is only 0.04\% lower than the peak accuracy in the baseline ST encoder. It is also a 1.1\% improvement over the peak accuracy of 97.8\% reported in the reference encoder \cite{rahimi_emg}, which uses the binding result of two channels to break ties while performing the associative memory search using Hamming distance.

As the \textit{N}-gram size decreases, the spatial encoding plays a more prominent role than the temporal encoding. Thus, the higher accuracy delivered by the in-memory ST encoder compared to the conventional ST encoder (see Fig.~\ref{fig:accuracy}(a)) for smaller \textit{N}-gram sizes can be explained by the spatial bundling operation being relocated downstream in the in-memory ST encoder. This allows retaining more useful spatial information in the encoded \textit{N}-gram hypervector.

Fig.~\ref{fig:accuracy}(b) shows that the in-memory encoder simulated with the statistical model exhibits an increasing accuracy drop, compared to the same encoder running in software, as the quantization levels increase. 
This is because as the quantization levels are increased, the PCM crossbar array size increases linearly, thereby amplifying the negative effect of spatial PCM variations on the classification accuracy. 
However given that the in-memory ST encoding operations in Equations \eqref{eq:channelbound} to \eqref{eq:majority_2} involve element-wise operations, or operations that involve neighboring elements, the crossbar array can be easily split into several realistic size\cite{le2018mixed} subarrays with simple single wire connectivity between subarray peripherals. This facilitates the silicon realization with negligible additional cost in terms of energy and area, as well as the mitigation of the effect of PCM spatial variations by compensating for subarray level conductance variation.


\subsection{Energy efficiency study and benchmark}
\label{sec:energy}
We performed an energy efficiency study of the in-memory ST encoder. The binary PCM device specifications given in \cite{Karunaratne2020} are used as the reference for obtaining power and timing numbers of the crossbars. The power and timing numbers for the digital peripherals are obtained from component-wise simulations of a post-synthesis netlist generated with 65nm CMOS technology. For comparison, we considered an equivalent digital ST encoder operating entirely in CMOS, whose power and timing numbers are obtained from a component-wise simulation of the post-synthesis netlist generated at the same technology node. Both digital peripherals and the equivalent CMOS encoder operate at  440\;MHz and 1.2\;V supply voltage.
\begin{figure}[tb]
\centering
\includegraphics[width=1\columnwidth]{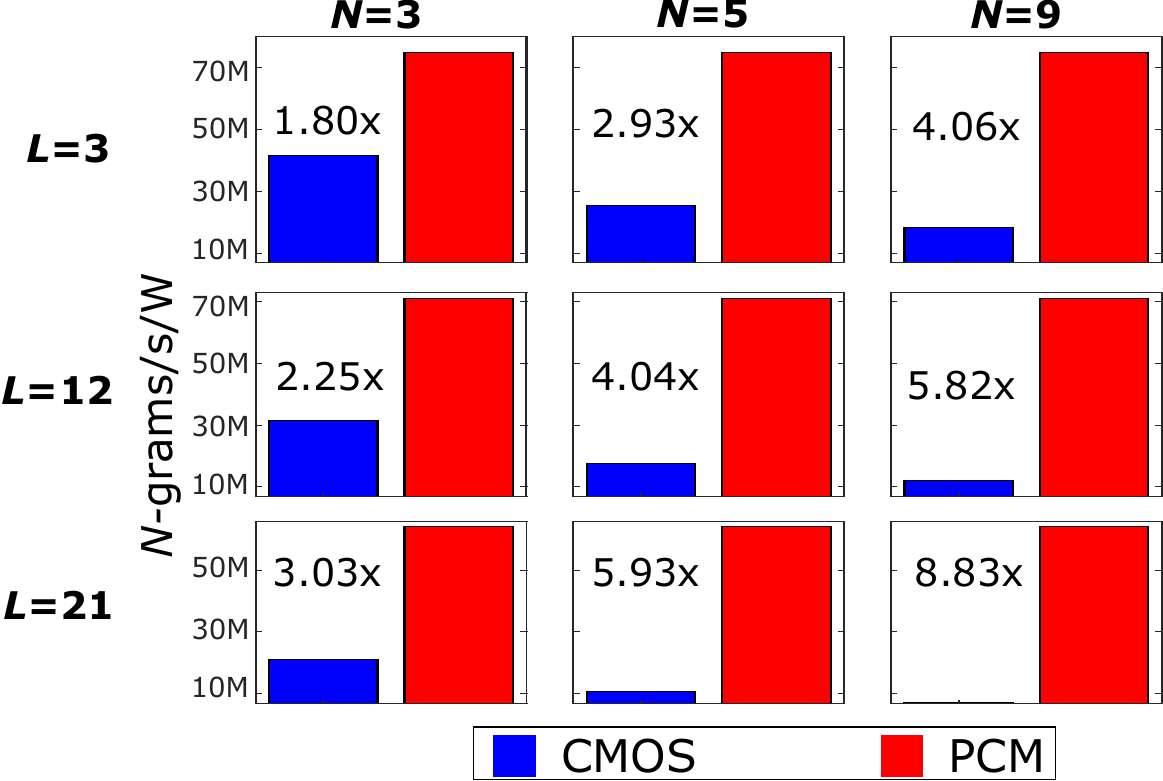}
\caption{Energy efficiency comparison, measured in terms of number of \textit{N}-grams encoded per second per Watt, between the in-memory ST encoder vs a dedicated digital CMOS ST encoder, on 9 different pairs of \textit{N}-gram size and quantization level parameter combinations. The energy efficiency gain of the in-memory ST encoder compared to digital ST encoder for each parameter combination is provided within the corresponding cell.}
\label{fig:energy}
\end{figure}

We observe that the in-memory ST encoder is able to produce 31.5M, 18.9M, and 10.5M \textit{N}-grams/s for \textit{N}-gram sizes of 3, 5, and 9, respectively, which is a throughput improvement of 9.74$\times$  over the digital counterpart and a 0.28M$\times$ improvement over the 10\,ms fixed latency PULP-HD \cite{pulphd}. The total area of the in-memory ST encoder varies from 0.37 to 0.44 and 0.51\;$mm^2$ when quantization levels are set to 3, 12 and 21, respectively. This is a 3.36$\times$, 5.46$\times$, and 6.97$\times$ area reduction, respectively, compared to the digital CMOS ST encoder. When breaking down the area numbers further we find that, irrespective of the number of quantization levels, a fixed area of 0.32\;$mm^2$ is occupied by the digital peripheral logic including the circular buffer, the binder and the bundler; another area of 0.02\;$mm^2$ is occupied by the row decoders and the sense amplifiers; while the rest of the area is taken by the PCM device array itself.

We estimated the energy efficiency of the in-memory ST encoder and compared it with the digital ST encoder as shown in Fig.~\ref{fig:energy}. The in-memory ST encoder achieves a peak energy efficiency of 75.1M\;\textit{N}-grams/s/W with \textit{N}=3 and \textit{L}=3. The total energy required for encoding an \textit{N}-gram using this configuration is 13.3\,nJ, 91.4\% of which are spent on digital peripheral circuits, 8.4\% on sense amplifiers/row decoders, and a mere 0.12\% on PCM devices.
This results in a 1.80$\times$ energy efficiency gain compared to a similarly configured digital ST encoder. The gain improves to a maximum of 8.83$\times$ as the \textit{N}-gram sizes and quantization levels are increased (see Fig.~\ref{fig:energy}). 

Table~\ref{tab:comparison} presents physical and performance characteristics of 1-core and 4-core PULP-HD encoders compared with a dedicated digital CMOS encoder and the proposed PCM-based in-memory ST encoder with the same parameter configuration ($N=3$, $L=21$). The energy required for \textit{N}-gram encoding is reduced from the $\mu$J range for the PULP-HD implementations to the $nJ$ range for the dedicated CMOS and PCM-based in-memory encoders. In summary, when compared with 1-core and 4-core PULP-HD encoders, the in-memory ST encoder achieves 1320$\times$ and 284$\times$ higher energy efficiency, respectively.

\begin{table}[]
\caption{Encoder Performance Comparison}
\label{tab:comparison}
\begin{tabular}{lrrrr}
\multicolumn{1}{c}{{ }} & \multicolumn{1}{c}{{ \textbf{\begin{tabular}[c]{@{}c@{}}PULPv3\\ 1 core\end{tabular}}}} & \multicolumn{1}{c}{{ \textbf{\begin{tabular}[c]{@{}c@{}}PULPv3\\ 4 cores\end{tabular}}}} & \multicolumn{1}{c}{{ \textbf{CMOS}}} & \multicolumn{1}{c}{{ \textbf{PCM}}} \\
\hline
{ Technology (nm)}      & { 28}                                                                                   & { 28}                                                                                    & { 65$^*$}             & { 65$^*$}            \\
{ Supply Voltage (V)}   & { 0.7}                                                                                  & { 0.5}                                                                                   & { 1.2$^*$}                              & { 1.2$^*$}                             \\
{ Frequency (Hz)}       & { 53.3M}                                                                                & { 14.3M}                                                                                 & { 440M$^*$}                             & { 440M$^*$}                            \\
{ Throughput (Ngram/s)} & { 108}                                                                                  & { 111}                                                                                   & { 3.23M\enspace}                             & { 31.5M\enspace}                            \\
{ Core Power (mW)}      & { 1.90}                                                                                 & { 0.42}                                                                                  & { 175$^*$}                              & { -\enspace}                                \\
{ Energy (J/Ngram)}    & { 17.5$\mu$}                                                                                & { 3.79$\mu$}                                                                                 & { 54.1n\enspace}                              & { 13.3n\enspace\vspace{0.1cm}}   \\
\hline
\multicolumn{5}{l}{\scriptsize  $^*$fast clock CMOS components}  
\end{tabular}
\end{table}


\section{Conclusion}
In this paper, we have demonstrated HDC encoding on spatio-temporal signals using in-memory computing techniques on memristive crossbar arrays. This approach allows selecting separate parameter combinations for each channel, further enhancing the flexibility of the encoding process. By simulating our architecture with a phase-change memory statistical model, we obtain a peak classification accuracy of 98.9\% (within 0.04\% of the baseline), while achieving 1.80$\times$--8.83$\times$ higher energy efficiency over a dedicated digital CMOS encoder and a 284$\times$ gain energy efficiency over an encoder running on a low-power general purpose computing platform.

\section*{Acknowledgment}
This work was supported in part by the European Research Council through the European Union’s Horizon 2020 Research and
Innovation Program under Grant 682675 and in part by the European Union’s Horizon 2020 Research and Innovation Program
through the project MNEMOSENE under Grant 780215.

\tiny

\end{document}